# Long-lived discrete breathers in free-standing graphene

Alberto Fraile[1], Emmanuel N. Koukaras[2], Konstantinos Papagelis[2], Nikos Lazarides[1,3] and Giorgos P. Tsironis[1,3]

[1]*Crete Center for Quantum Complexity and Nanotechnology, Department of Physics, University of Crete, P. O. Box 2208, 71003, Heraklion, Greece*
[2]*Institute of Chemical Engineering and High Temperature Chemical Processes (ICE/HT), Foundation for Research and Technology-Hellas (FORTH), P.O. Box 1414, GR-265 00 Patras, Greece.*
[3]*Institute of Electronic Structure and Laser, Foundation for Research and Technology-Hellas, P.O. Box 1527, 71110 Heraklion, Greece*

Intrinsic localized modes or discrete breathers are investigated by molecular dynamics simulations in free-standing graphene. Discrete breathers are generated either through thermal quenching of the graphene lattice or by proper initialization, with frequencies and lifetimes sensitively depending on the interatomic potential describing the carbon-carbon interaction. In the most realistic scenario, for which temperature-dependent molecular dynamics simulations in three dimension using a graphene-specific interatomic potential are performed, the breather lifetimes increase to hundreds of picoseconds even at relatively high temperatures. These lifetimes are much higher than those anticipated from earlier calculations, and may enable direct breather observation in Raman spectroscopy experiments.

Recent technological achievements have allowed for the isolation of single graphene (GE) sheets either by chemical exfoliation of bulk graphite [1, 2] or by epitaxial growth on metal substrates through thermal decomposition of SiC [3]. As a truly two-dimensional (2D) system, it provides a framework for new type of electronic and magnetic devices [4]. While the electronic properties of GE have been exhaustively investigated [5], its mechanical and thermal properties are not quite thoroughly analyzed. In particular, while a reliable linear phonon spectrum can be obtained numerically and compared with the experimental one, much less is known about the nonlinear vibrational modes in GE. Since all potentials modeling vibrational properties of GE are nonlinear, it is natural to expect that intrinsic localized modes, i.e. discrete breathers (DBs), may be formed. DBs are spatially localized and time-periodic vibrational modes that form spontaneously in nonlinear lattices [6-8]; they have been assessed experimentally in several systems, including solid state mixed-valence transition metal complexes [9], quasi-one dimensional antiferromagnetic chains [10], micromechanical oscillators [11], optical waveguide systems [12], proteins [13], and NaI crystals [14].

A reliable computational prediction of possible DB excitations in GE, their lifetime and spectral features, will enable their direct experimental observation and facilitate the design of efficient future devices at relatively high temperatures. In the present work we perform extensive molecular dynamics (MD) simulations that reveal the existence of DBs in free-standing GE both at zero temperature in 2D as well as at finite-temperatures in three dimensions (3D). To this end, we resort to Sandia National Laboratories Large-Scale Atomistic/ Molecular Massively Parallel Simulator (LAMMPS) [15]. In order to demonstrate the DB onset and stability range we have tested several interatomic potentials (IPs) that have been utilized in the past to model carbon systems (Tersoff [16], AIREBO [17], LCBOP [18], CBOP [19], among others). However, we eventually focus on a graphene-specific Tersoff potential [16], hereafter referred to as Tersoff'10, which is of the form

$$E_{ij} = V_{ij}^R + b_{ij}V_{ij}^A \qquad (1)$$

The functions *VR(r)* and *VA(r)* are pair-additive interactions that represent all interatomic repulsions (core–core, etc.) and attraction from valence electrons, respectively. The quantity *rij* is the distance between pairs of nearest-neighboring atoms *i* and *j* and *bij* is a bond order between atoms *i* and *j*. As it would be expected, the stability and lifetimes of DBs in GE is strongly potential-dependent [20-23]. AIREBO and LCBOP potentials include angular, dihedral (out of plane distortions), as well as long-range terms. The comparison of the results obtained using AIREBO and LCBOP points to a destructive effect of the long-range part of the potentials on the stability of the DBs. This is more evident when LCBOP is compared with the CBOP. The Tersoff'10 potential is a reparametrized version of the original Tersoff'89 IP that provides much better agreement with the experimental phonon velocities and frequencies, without significantly altering the agreement to other structural data. The choice of that potential is based on its performance on particular features regarding the vibrational properties of GE. Along with LCBOP, the Tersoff'10 IP provides the most accurate overall description for its phonon dispersion curves. In particular, the Tersoff'10 IP produces more accurate LA and ZA branches but less accurate ZO and TA branches than LCBOP at temperature T = 300 K. Moreover, it is the only one to produce a linear temperature-dependence of the doubly degenerate Raman active $E_{2g}$ mode of the Γ point [24].

We consider these two features important for the correct simulation of DBs in free-standing GE.

The temperature-dependence of the vibrational GE spectra was investigated using the original Tersoff'89, Tersoff'10, LCBOP, and AIREBO potentials. The MD simulations were performed using a periodic triclinic computational cell of 20x20 unit cells (overall 800 carbon atoms). The computational cell was relaxed for each potential and the corresponding lattice parameter was calculated and used in defining the BZ edges in each case. A very fine time-step of 0.05 fs was used and the trajectory and velocities were saved every 10 time-steps. These simulations typically run for about 32.8 ps each (655360 time-steps). For the Tersoff'10 potential, the GE vibrational response was analyzed by producing the dispersion curves at temperatures T=60 K, 500 K, and 1500 K [24]. For that IP, the strongest temperature-dependence is observed on the optical branches. Upon increasing of temperature, the frequencies lower by as much as 64 cm–1 (~2 THz).

Discrete breathers in GE may be generated by either local initial displacements (and subsequent MD evolution), or by thermal quenching of the GE lattice. The former method consists of displacing a few atoms deeply in the GE lattice according to an approximate solution obtained by the rotating wave approximation of the equations of motion (Figure 1) [25]. The simulations were carried out in the microcanonical ensemble (NVE) in 2D using the Tersoff'10 potential with a time-step $\Delta t=10^{-5}$ ps and periodic boundary conditions. The GE samples typically contain 15000 atoms, so that edge effects do not affect DB stability.

Finite size effects have been investigated by performing repeated simulations using computational cells of different sizes. The initial displacement, d, ranges from 0.1 to 0.3 Å. Depending on the value of d, two different DB configurations, say Type-1 and Type-2, have been observed. Type-1 breathers are observed for d in the range 0.15-0.19 Å (Figure 2a), while type-2 DBs are observed for d in the range 0.27-0.3 Å

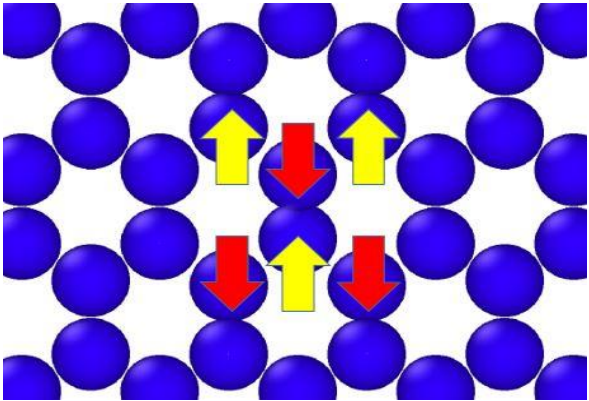

**FIG. 1.** (Colour online) Schematic illustration of the initial condition to generate discrete breathers (Arrows are not at scale).

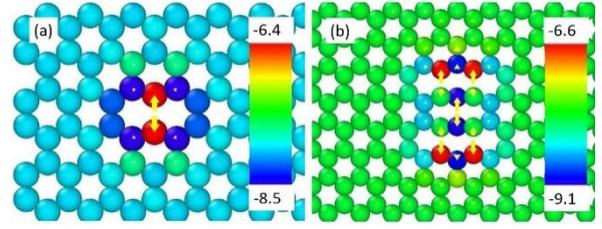

**FIG. 2.** (Colour online) Type-1 (a) and Type-2 (b) breathers in a single graphene sheet at T=10 K. Color scale correspond to the total energy in eV/atom. The yellow arrows show the displacements with respect to the equilibrium position after stabilization. The length of the arrows is 3 times the real value for the shake of clarity.

(Figure 2b). Both Type-1 and Type-2 DBs exhibit almost steady evolution for relatively long times. The *central* atoms of Type-1 DBs (red atoms in Figure 2a) execute large amplitude periodic oscillations while the outer ones (blue atoms in Figure 2a) small amplitude oscillations. Typical oscillation patterns of the inner (central) and outer atoms of a Type-1 DB are shown in Figure 3 (Supporting Information Video 1 [26]). The phases of large and small amplitude oscillations differ by π in that DB configuration. The frequency of the oscillations of the inner atoms here is not constant with time but increases up to twice its value at the same time that the amplitude of the oscillations clearly decreases (after 20 ps).

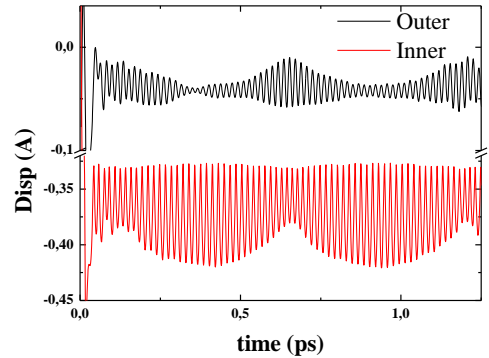

**FIG. 3.** (Colour online) Displacement in the *y* direction (in Å) of inner (red line) and outer (black line) atoms in a Type-1 breather. Notice the break in the y-axis. Both oscillations correspond to a frequency ~56 THz.

The configuration of Type-2 DBs is much more complex and only some of the atoms exhibit regular oscillations (Supporting Information Video 2 [26]). In both types of DBs however the energy remains almost constant for long times, during which their frequency is calculated to be ~56 THz, i.e., above the highest frequency of the upper optical phonon dispersion branch. The total energy stored in the central DB atoms is considerably higher than that of the outer atoms, around 0.1 eV/atom for both DB types. That energy slowly decays with time as can be seen in Figure 4. From an extrapolation of the energy decay, we can estimate the lifetime of Type-1 and Type-2 DBs to be 500 ps and 450 ps, respectively. The properties of Type-1 and Type-2 DBs obtained from the NVE simulations in 2D using the Tersoff'10 potential are summarized in Table 1. The calculated DB lifetimes using the Tersoff'10 potential are much longer than those reported before, e.g., in finite-temperature simulations using the Brenner potential (~1 ps at T=10 K) [20] and in MD simulations using an AIREBO-type potential (~30 ps) in strained graphene [21]. The corresponding breather frequencies have been reported to be 47 THz and ~27- 32 THz, respectively.

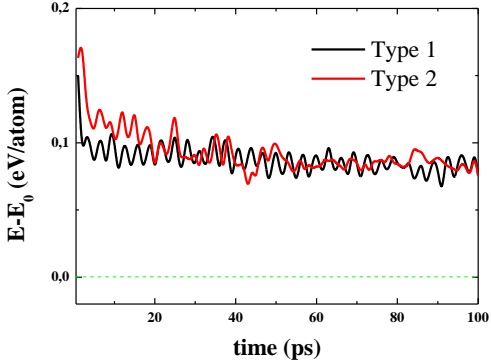

**FIG. 4.** (Colour online) Time evolution of the total energy per atom, of the two types of breathers, minus the background (green dashed line) energy $E_0$, in eV/atom.

| DB type | Properties | | | | |
|---|---|---|---|---|---|
| | d (Å) | ν (THz) | A (Å) | $E_{ex}$ (eV/at) | τ (ps) |
| 1 | 0.15-0.19 | 56 ± 0.4 | 0.11 | 0.1 | 500 ±50 |
| 2* | 0.27-031 | 59 ± 1 | 0.05 | 0.1 | 450 ±50 |

**Table 1**. Properties of Type-1 and Type-2 DBs observed in NVE simulations in 2D using the Tersoff'10 potential, where d is the initial displacement, ν and A the frequency and amplitude of the oscillations, respectively, τ the calculated lifetime, and $E_{ex}$ the excess energy stored in the DB. (*) The oscillation corresponds to the outer atoms in dark blue (Figure 2b).

We now turn our attention to the effect of temperature on the stability of the DBs. The GE sample is first equilibrated to the desired temperature for 10 ps. A DB is then created by applying suitable initial displacements, as described earlier, and the simulation continues keeping the temperature constant at the desired value. It is anticipated that the DB lifetime is temperature-dependent since thermal noise will eventually destroy its stability. The results presented so far were constrained in 2D for simplicity; that constraint, however, excludes the possibility of off-plane phenomena by completely quenching intrinsic thermal rippling [27-29]. For that reason, our finite-temperature NPT simulations at zero pressure are performed in 3D and reveal that both Type-1 and Type-2 DBs remain stable up to temperatures as high as 2000 K. Note that this is very near the melting temperature of GE for the Tersoff'10 potential, that is ~2100 K. The difference between the highest and lowest atomic displacements along the z-axis (perpendicular to the GE plane) in a computational cell consisting of 15000 atoms at high temperatures (from 1500 to 2000 K), resulting from the NPT simulations in 3D, is almost constant and around 15 ±2 Å for a cell length of 250 Å. That curvature is rather small and moreover it is in good agreement with the experimentally obtained ones [30]. For temperatures increasing from T=300 K to 2000 K, the curvature / corrugation increases but the DBs remain quite unaffected as far as their lifetime τ is concerned (Supporting Information Video 3 [26]). For small initial displacements of the order of 0.1 Å, the DBs are very sensitive to thermal vibrations; even at temperatures as low as 10 K, they can be easily destroyed. For larger displacements, the DBs are stabilized despite high temperatures, with recorded lifetimes longer than τ~500 ps. For example, if the displacement is 0.15 Å (0.2 Å), the Type-1 DB remains stable for long times if the temperature is lower than a critical one, $T_C$=500 K ($T_C$ =1900 K). At high temperatures (T> 1500 K), the DB structure is slightly distorted as compared to the symmetric configurations in Figure 2. However, some atoms still remain in a higher or a lower energy level for long times (more than 500 ps). The DBs created with initial displacements 0.25 (of Type-1) or 0.3 Å (of Type-2) can be regarded to be stable even at T= 2000 K. A typical Type-2 DB configuration is shown in Figure 5a with its 3D structure being tilted out of plane. Table 2 summarizes the results of this paragraph.

The value of the critical temperature $T_C$ clearly depends on the initial displacement used to generate a DB, as well as the type of the DB; for $T > T_C$, however, both types of DBs rapidly decay. Generally speaking, the $T_C$ of a particular DB increases with increasing the initial displacement, d, indicating that more energetic breathers survive for longer times. Importantly, at T > 1600 K, thermal excitation of DBs has been observed (Supporting Information Video 4 [26]). Notably, thermal excitation of DBs that arise from anharmnonicity, that is expected to be strong at high temperatures, has been also demonstrated in ionic perfect crystals [31].

| Disp (Å) | 0.1 | 0.15 | 0.2 | 0.25 | 0.3 |
|---|---|---|---|---|---|
| $T_C$ (K) | 10 | 500 | 1900 | 2000 | 2000 |
| Lifetime (ps) | < 0.5ps | > 500 | > 500 | > 500 | > 500 |

**Table 2**. Breather lifetimes at high temperatures as a function of the initial displacement. When the temperature is lower than $T_C$ the lifetime is at least the one quoted in the second row. For all displacements but d=0.3 Å the DBs are of Type-1, while for d=0.3 Å the breathers are of Type-2.

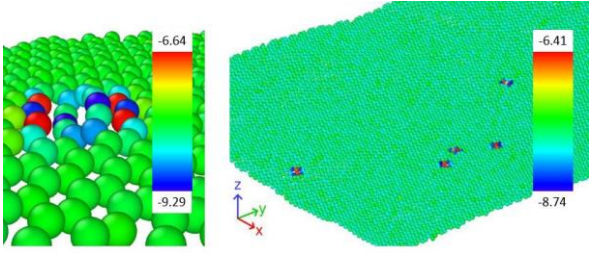

**FIG. 5**. (Colour online) (a) Type-2 DB in a graphene sheet heated at 1700 K in 3D. The breather structure is tilted out of plane; (b) Type-1 breathers in a graphene sheet after quenching from 1600 K. The rainbow scales represent the total energy in eV.

Temperature effects upon DB lifetimes can be better understood by analysing the corresponding effects on the phonon dispersion curves. The frequency of the LO/TO modes slightly decreases with increasing temperature with a rate very closed to that experimentally observed [32]. At the same time, the DB oscillation frequency (of Type-1 here), or more precisely the oscillation of the inner (central) atoms shown in red colour in Figure 2a, decreases faster and drops below that of the LO/TO modes at temperatures higher than 400 K. At such temperatures, the interaction with the (linear) LO/TO phonon modes removes energy from the DB which slowly decays (within a few hundreds of ps).

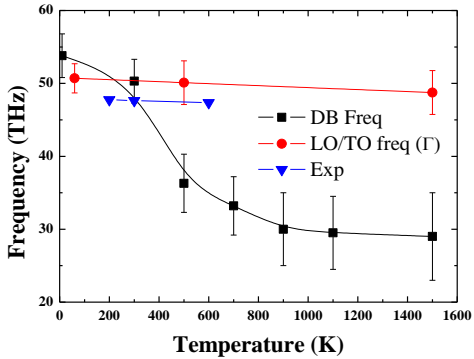

**FIG. 6**. (Colour online) Type-1 breather frequency (black squares) and LO/TO frequency modes at Γ point (red circles) for different temperatures. Experimental values from [27].

A better suited for experiments and perhaps more natural method for DB generation is that of thermal quenching [33, 34]. First, a GE sample is heated up to a particular high temperature $T_h$ above the Debye temperature in which all the phonon modes are populated. At that temperature, typically at $T_h > 1600$ K (in references [35, 36] the values 2100 K and 2300 K are quoted, in reasonable agreement), the system is equilibrated for $t_{eq}$ ps in the NPT (isothermal–isobaric) ensemble, and subsequently it is cooled down rapidly (quenching) to a relatively low temperature $T_f$. In our simulations, $T_f$ ranges from 100 K to 500 K, while the effective quenching time is about 1-2 ps that corresponds to quenching rates from 2000 K/ps to 1000 K/ps. In accordance with our results on the temperature-dependent lifetime of DBs, the number of DBs generated by thermal quenching does not depend significantly on the final temperature $T_f$ as long as this is around room temperature.

However, the number of DBs at $T_f$ does depend on the temperature $T_h$ as well as the equilibration time $t_{eq}$ at that temperature. Indeed, the number of DBs at $T_f$ increases linearly with the equilibration time $t_{eq}$ at $T_h$, while it increases exponentially with $T_h$. The maximum density of DBs obtained by quenching the samples from 2000 K is about 2-4 per thousand atoms, distributed randomly in the sample (Figure 5b). The interaction between DBs is weak, and each of them keeps its identity even though they may approach each other at distances less than 10 Å.

In summary, the existence of long-lived DBs in a single GE layer is demonstrated by MD simulations using the graphene-specific potential Tersoff'10. The latter describes accurately the phonon dispersion curves, especially the optical branches, as well as their temperature-dependence. Moreover, it reproduces the linear temperature-dependence of the doubly degenerate Raman active $E_{2g}$ mode of the Γ point [24]. These features make the Tersoff'10 the most appropriate for the search of DBs in GE and the reliable determination of their lifetimes and oscillation frequencies. Our work does not merely provide additional evidence that DBs are likely to exist in GE, but it makes lifetime predictions of the order of hundreds of picoseconds, much longer than those reported so far. Moreover, these lifetimes of GE breathers persist when dimensionality increases from 2D to 3D. In 3D, the type-2 DBs abandon the symmetric configurations observed in 2D, and they acquire a 3D structure that is tilted with respect to the GE plane. The distorted 3D breathers are fairly stable with hundreds of picoseconds lifetimes at high temperatures even close to the melting point. A non-negligible density of DBs have been also obtained by thermal quenching in NPT simulations in 3D at zero pressure. In this case we only obtain Type-1 breathers, implying that the thermal energy is not enough to excite Type-2 DBs. Higher temperatures result in a proliferation of defects and eventual melting. Our simulations provide a clear estimation for the range of temperatures in which DBs are expected to be thermally excited (above 1500 K and below 2000 K), as well as the dependencies of their number density. The present work represents a step forward for understanding the nonlinear physics of GE with a possible impact in graphene-based future technological applications.

**Acknowledgments**

This work was partially supported by the European Union's Seventh Framework Programme (FP7-REGPOT-2012-2013-1) under grant agreement no 316165.